\documentclass[aps,pra,longbibliography=false,twocolumn]{revtex4-2}
\usepackage{amsfonts}
\usepackage{amsmath}
\usepackage{amssymb}
\usepackage{ragged2e}  
\usepackage{subfig}
\usepackage{braket}
\usepackage{graphicx} 
\graphicspath{ {images/} }
\usepackage[dvipsnames]{xcolor}

\usepackage{hyperref}
\hypersetup{
    colorlinks=true,
    linkcolor=blue,
    citecolor=blue,      
    urlcolor=blue,
    }

\newcommand{\pasqal}{Pasqal, 7 rue Léonard de Vinci, 91300, Massy, France}
\newcommand{\panasonic}{Panasonic Industry Co., Ltd., 1006, Oaza-Kadoma, Kadoma, Osaka, Japan}

\newcommand{\<}{\langle}
\renewcommand{\>}{\rangle}

\makeatletter
\renewcommand{\@makecaption}[2]{%
  \vskip\abovecaptionskip
  \justifying\small\textbf{#1.} #2\par
  \vskip\belowcaptionskip
}
\makeatother

\begin{document}

\title{ Materials Discovery With Quantum-Enhanced Machine Learning Algorithms}

\author{Ignacio F. Graña}
\affiliation{\pasqal}
\author{Savvas Varsamopoulos}
\affiliation{\pasqal}
\author{Tatsuhito Ando}
\affiliation{\panasonic}
\author{Hiroyuki Maeshima}
\affiliation{\panasonic}
\author{Nobuyuki N. Matsuzawa}
\affiliation{\panasonic}

\begin{abstract}
Materials discovery is a computationally intensive process that requires exploring vast chemical spaces to identify promising candidates with desirable properties.
In this work, we propose using quantum-enhanced machine learning algorithms following the extremal learning framework to predict novel heteroacene structures with low hole reorganization energy $\lambda$, a key property for organic semiconductors.
We leverage chemical data generated in a previous large-scale virtual screening \cite{Matsuzawa2020} to construct three initial training datasets containing 54, 99 and 119 molecules encoded using $N=7,16$ and 22 bits, respectively.
Furthermore, a sequential learning process is employed to augment the initial training data with compounds predicted by the algorithms through iterative retraining.
Both algorithms are able to successfully extrapolate to heteroacene structures with lower $\lambda$ than in the initial dataset, demonstrating good generalization capabilities even when the amount of initial data is limited.
We observe an improvement in the quality of the predicted compounds as the number of encoding bits $N$ increases, which offers an exciting prospect for applying the algorithms to richer chemical spaces that require larger values of $N$ and hence, in perspective, larger quantum circuits to deploy the proposed quantum-enhanced protocols.
\end{abstract}

\maketitle

\section{Introduction}

The discovery of novel materials represents a central challenge in materials science with many industrially relevant applications. Traditional experimental and computational approaches are often inefficient and consume large amounts of resources navigating the vast chemical design space. This typically results in low success rates and extensive development timelines, often stretching over many years \cite{LIU2017159, rajan2005materials}. Consequently, there is a pressing need for novel methods that can address these limitations by exploring the design space more efficiently.

In this work we focus on computational methods for the discovery of novel organic semiconductors, a family of chemical structures that has recently drawn great attention. 
These compounds are known to offer a variety of interesting properties such as being lightweight, requiring simple processing and having low cost of manufacturing.
Various classes of molecules, such as oligoacenes \cite{Anthony2006},  polythiophenes \cite{Perepichka2005},  heteroarenes \cite{Kang2011} and fullerenes \cite{Sieval2024} have been investigated as a novel organic semiconductor compound. 
To this date, the highest electron and hole mobilities reported are 11 $\rm{cm}^2$/(V s) for $\rm{C}_{60}$ \cite{Li2012} and 40 cm$^2$/(V s) for rubrene \cite{Takeya2007}, respectively. 
However, these mobility values are still significantly lower than those of silicon, which has mobilities of 1200 $\rm{cm}^2$/(V s) for electrons and 500 $\rm{cm}^2$/(V s) for holes \cite{Prince1954}. 
Therefore, the applications of such compounds are currently limited by their low carrier mobilities. Estimating this key property is often computationally too expensive, so other metrics such as the reorganization energy $\lambda$ have been identified as good estimators of the charge mobility while remaining computationally tractable. It should be noted that based on Marcus theory \cite{Marcus1956,Marcus1957a,Marcus1957b,Marcus1965}, it is expected that carrier mobilities tend to become larger when $\lambda$ becomes smaller. 

Virtual screenings calculating these figures of merit for a large body of candidates have been commonly used to identify new promising compounds, despite being computationally intensive due to the large solution space. For example, Schober et al. \cite{Schober2016} calculated reorganization energy and intermolecular electronic coupling of 95,445 molecular crystal structures taken from the Cambridge Structural Database \cite{Allen2002} and succeeded in extracting four promising molecules. A massive theoretical screen of hole conducting heteroacene molecules has also been reported, where hole reorganization energies of quarter million heteroacenes were calculated by using a cloud computing environment and various promising structures were identified \cite{Matsuzawa2020}. 

Recently, machine learning models have been proposed as reliable predictors of target molecular properties for exploring novel organic semiconducting molecules \cite{Marques2021, Kunkel2019, Nagasawa2018, Antono2020, Ando2022, Staker2022}, potentially reducing the computational requirements of data-hungry virtual screenings. For example, a database of more than 64,000 organic molecular crystals was used to design organic semiconductors applying a data-mining approach \cite{Kunkel2019}. In another study, conjugated semiconducting polymers for organic solar cells were screened based on a machine learning model made using more than 1000 experimental values \cite{Nagasawa2018}. A sequential learning approach based on Bayesian optimization and machine learned models in combination with molecular dynamics and density functional theory (DFT) calculations were performed to explore molecules exhibiting improved mobility and $\lambda$ \cite{Antono2020, Ando2022}. 
Finally, various generative artificial intelligence methods have also been applied to design molecules with improved $\lambda$ by using the aforementioned quarter million DFT calculated reorganization energies as a training dataset for machine learning \cite{Marques2021, Staker2022}.

In parallel to these ML advances, quantum computing has emerged as an alternative technology with the potential to speed up some of the bottlenecks in computational materials design \cite{Clinton_2024,Babbush_2018}. 
Combined approaches leveraging existing classical high-performance computing and quantum processing resources offer exciting avenues for more efficient algorithmic pipelines \cite{ALEXEEV2024666}.
Within the field of quantum algorithms, Quantum Machine Learning (QML) represents a new paradigm that may provide certain advantages compared to conventional ML algorithms, such as greater expressibility and generalization capabilities \cite{schuld2021effect, Caro_2022}.

In this work, we carry an initial study of the potential of \textit{quantum-enhanced} ML algorithms in the field of materials discovery by focusing on the specific problem of predicting low-$\lambda$ organic semiconductors. 
Due to the large chemical design space, we focus our search on fused heteroacene structures containing sulfur heteroatoms, which are promising candidates for high hole mobility, where intensive experimental studies have been performed aiming to identify chemical structures of heteroacenes that exhibit as high hole mobility as possible \cite{Takimiya2011, Shinamura2010, Okamoto2013, Okamoto2018, Haas2009}. To benchmark the algorithms, three different training datasets were constructed by partially leveraging data previously generated in a large-scale virtual screening of heteroacene structures \cite{Matsuzawa2020}.
We encoded the chosen heteroacene structures via a binary representation using $N$ bits to encode information about its molecular structure.
Three datasets were built using $N=7,16$ and 22 bits to map the structures.
The first dataset used $N=7$ to encode heteroacene structures incorporating 4 rings.
The larger datasets using $N=16$ and $N=22$ bits to represent 8-ringed heteroacenes were employed to test the algorithms with a larger solution space.
With the $N=16$ and $N=22$ dataset, a sequential learning process \cite{ rohr2020benchmarking, Palizhati2022, volker2021sequential, Balachandran2016} was carried to iteratively expand the training data according to the optimal compounds predicted by the algorithms. 
The algorithms were retrained with the augmented training data at each iteration, incrementally improving the quality of the predictions while biasing the algorithms towards the low-$\lambda$ regime.

We employ algorithms based on the \textit{extremal learning} framework proposed in \cite{patel2021extremallearningextremizingoutput}. 
The aim is to find the inputs that \textit{extremize} the figure of merit of a given dataset. 
In our case, the input would be a chemical structure and the figure of merit the hole reorganization energy, which we would like to minimize. 
Extremal learning algorithms do so via two distinct phases. 
First, in the \textit{learning phase}, a ML model is trained to learn the relation between the inputs and the figure of merit, potentially from few data.
In the \textit{selection phase}, a procedure is followed to identify the optimal (extremal) inputs based on the learned model. 
The outcome of the algorithm is a list of candidate inputs with a high probability of being close to the optimal solutions, based on the chosen figure of merit. 
Depending on the particular algorithm, both the learning and the selection phase can be done via either classical or quantum computation. 

Two quantum-enhanced extremal learning algorithms are studied here, namely \textit{Quantum Extremal Learning} (QEL) \cite{Varsamopoulos2024} and the \textit{Factorization Machine + Quantum Approximate Optimization Algorithm} (FM+QAOA) \cite{Kitai2020}.
The variant of QEL used here employs a QML model (Quantum Neural Network) in the learning phase and a classical method in the selection phase. 
On the other hand, FM+QAOA follows the opposite approach, using a classical ML model (a Factorization Machine) in the learning phase and a QML model in the selection phase (QAOA).
We numerically benchmark FM+QAOA with the three molecular datasets previously described, while QEL was only benchmarked with the $N=7$ and $N=16$ datasets due to the large computational cost of emulating it in classical hardware. 
Both algorithms are shown to successfully  generalize to low-$\lambda$ compounds with surprisingly few data.
Interestingly, we observed that the performance of both algorithms improves as the number of bits $N$ in the encoding increases. 
In particular, the amount of training data relative to the solution space required to generalize to low-$\lambda$ compounds decreases with increasing $N$. 
This offers an exciting outlook for the proposed extremal learning algorithms to be applied to larger and more complex datasets. 

The novel contributions of this work can be summarized as follows:
\begin{enumerate}
    \item An algorithmic pipeline combining classical and quantum resources is presented and applied to solve a non-trivial chemical task.
    \item We benchmark the quantum-enhanced algorithms with three different datasets constructed with real chemical data generated in a previous virtual screening. 
    \item We show how to encode the hetereocene structures via a binary encoding that can be naturally employed to upload the chemical data into a quantum circuit. 
    \item An iterative sequential learning process is carried, in which the initial datasets are augmented with the candidate compounds predicted by the algorithms.
    \item We show that both algorithms are able to successfully predict compounds with a lower hole reorganization energy than in the initial training datasets.
\end{enumerate}

The rest of the paper is structured as follows. 
In Section \ref{sec:problem_formulation}
we formalize the molecular discovery task under evaluation and present the dataset utilized to train the algorithms. 
Section \ref{sec:algorithmic_framework} describes the algorithmic framework and the sequential learning process. 
The numerical results are presented and discussed in Section \ref{sec:results} and finally, Section \ref{sec:conclusions} offers our conclusions and a summary of the work.

\section{Problem formulation}
\label{sec:problem_formulation}

In this work, we focus on the problem of predicting novel organic semiconducting molecular structures with a low hole reorganization energy, which we denote by $\lambda$ and measure in electronvolts (eV). We use this metric because it is considered to be a good estimator of the carrier mobility of the semiconductor. The carrier transport phenomena of molecules can be described based on Marcus theory \cite{Marcus1956, Marcus1957a, Marcus1957b}, where the charge-hopping rate of the local dimer of molecules, k, is formulated by Eq. (\ref{eq:marcus_equation}):

\begin{equation}
\label{eq:marcus_equation}
    k = \frac{2\pi}{\hbar} \left( \frac{H^2_{ab}}{\sqrt{4\pi\lambda k_B T}}\right) \text{exp}\left(\frac{-(\Delta G + \lambda)^2}{4\pi\lambda k_B T}\right),
\end{equation}

where $\hbar$ is Planck’s constant divided by $2\pi$, $k_B$ is Boltzmann’s constant, $T$ is the temperature, $\Delta G$ is the free energy difference for charge transfer, $\lambda$ is the reorganization energy, and $H_{ab}$ is the intermolecular electronic coupling. The four parameters $H_{ab}$, $\Delta G$, $\hbar$, and $T$ in this equation determine the charge transfer rate. Among them, $\lambda$ has a considerable effect on charge transfer as it decreases the transfer rate exponentially.

To construct the different training datasets used throughout this work, hole reorganization energies of selected heteroacenes were either calculated by applying Density Functional Theory (DFT) methods \cite{Kohn1965, Hohenberg1964} or taken from the previously calculated DFT results for the quarter million heteroacenes \cite{Matsuzawa2020}. 
The reorganization energy requires the calculation of four energy values: the optimized energy in the neutral state $E$, the energy of the neutral state in ion geometry $E^*$, the optimized energy in the ionic state $E_+$ and the energy of the ionic state in neutral geometry $E_+^*$. By using these quantities, $\lambda$ was estimated by applying Eq. \ref{eq:reorganization_energy} \cite{Deng2004}:

\begin{equation}
\label{eq:reorganization_energy}
\lambda = (E_+^* - E_+) + (E^* - E)	.
\end{equation}

The DFT results of $\lambda$ presented in this study were obtained using Materials Science Suite (Version 2021-3) \cite{Schrodinger2021} provided by Schro\"{o}dinger, LLC. 
Data analysis and interactive property visualizations were all carried out using the built-in functionality of the Materials Science Suite. 
Hole reorganization energies were computed using the optoelectronics module within the Materials Science Suite by applying the B3LYP \cite{Becke1993, Lee1988} hybrid density functional and LACV3P** basis set. This basis is a triple-$\zeta$ basis set that employs the 6-311G** basis set \cite{Ditchfield1971} in combination with the Hay and Wadt’s effective core potential \cite{Hay1985_1, Hay1985_2, Hay1985_3}, and implemented in the Jaguar DFT package (Version 2021-3) \cite{Bochevarov2013, Jaguar2021} provided by Schr\"{o}dinger, LLC.

In our datasets, we limit the chemical space to heteroacene structures with 4 rings in the $N=7$ encoding and with 8 rings in the $N=16$ and $N=22$ encodings. 
In all cases, the rings can either have no heteroatom (benzene ring) or have an added sulfur (thiophene ring). 
However, not all heteroacene structures fulfilling the aforementioned conditions can be encoded with our approach.
The number of such structures that can be encoded in binary strings is limited by the number of bits used in the encoding.

\subsection{Binary encoding}
There are various ways to encode a chemical compound into a computer program. Standard approaches include molecular encodings such as the SMILES line notation \cite{Weininger1988}, the use of molecular descriptors \cite{Mauri2017} or chemical graph embeddings \cite{Wang2024}. 
Here we use a different approach: we map the set of chemical structures of interest, denoted $\mathcal{S}$, to binary strings of length $N$, denoted $\{0,1\}^N$. 
Each bit carries information about either i) if a certain ring contains a benzene or a thiophene molecule or ii) how two adjacent rings are fused together. 
As we will discuss later, this binary representation allows for a natural encoding of the classical data into a quantum circuit, mapping a bitstring $x$ into the corresponding quantum computational basis state $\ket{x}$.

The map between chemical structures and binary strings can be formally written as $g: \mathcal{S}\rightarrow \{0,1\}^N$.
The specific choice of $g$ can greatly affect the performance of the overall algorithm. 
Note that $g$ cannot be a bijection as the cardinality of $S$ and $x\in\{0,1\}^N$ will in general not be equal.
However, it is important that all bitstrings $x\in\{0,1\}^N$ can be mapped back to a valid chemical structure to prevent the algorithm from predicting non-valid structures. 
This is, for every $x\in \{0,1\}^N$, there must be at least one chemical structure $s\in\mathcal{S}$ such that $g(s) = x$. Otherwise, an algorithm might generate candidates which do not correspond to valid structures, lowering the quality of the algorithmic predictions.
On the other hand, a single structure $s$ can correspond to more than one bitstring $x$, i.e. there might be more than one $x$ such that $x=g(s)$.
We refer to all bitstrings corresponding to the same chemical structure as \textit{duplicates}.
Duplicates are a direct consequence of the different cardinality of $S$ and $x\in\{0,1\}^N$.

\begin{figure}
    \centering
    \includegraphics[width=1\linewidth]{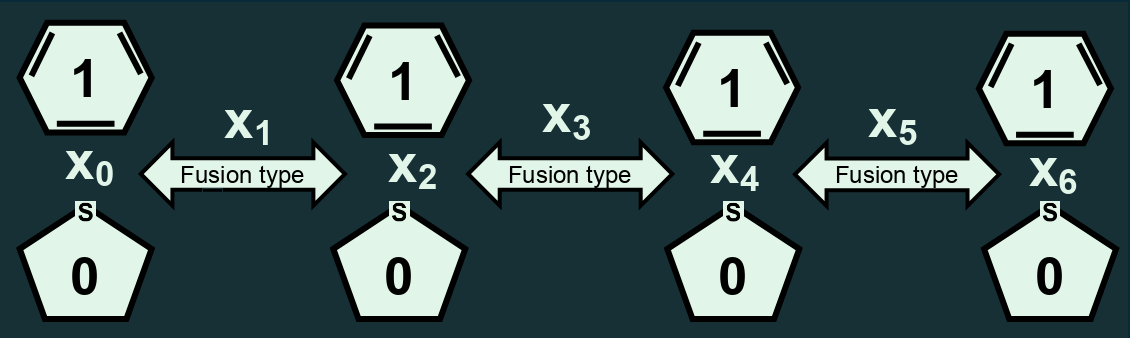}
    \caption{Diagram of the binary encoding used in the N=7 bits molecular dataset. Bits $x_0$, $x_2$, $x_4$ and $x_6$ are used to encode which heteroatom is in each ring, where a value of 0 means a sulfur is added (thiophene) and 1 means that no heteroatom is present (benzene). The rest of the bits $x_1$, $x_3$ and $x_5$ encode the orientation in which consecutive rings are fused to each other. }
    \label{fig:7_bits_encoding}
\end{figure}

\begin{table*}[ht]
\centering    \captionsetup{justification=centering}
    \caption{Definition of fused ring structure when a ring is added to a benzene ring for $N = 7$. }
    \begin{tabular}{cccccc}
        \hline
        $x_{n-3}$ & $x_{n-1}$ & $x_{n+1}$ & $x_n$ & $n = 1$ & $n = 3$ and $5$ \\
        \hline
        0 & 1 & 1 & 0 & naphthalene & naphtho[2,3-b]thiophene \\
        0 & 1 & 1 & 1 & naphthalene & naphtho[1,2-b]thiophene \\
        0 & 1 & 0 & 0 & benzo[b]thiophene & benzo[1,2-b:5,4-b']dithiophene \\
        0 & 1 & 0 & 1 & benzo[b]thiophene & benzo[1,2-b:4,5-b']dithiophene \\
        1 & 1 & 1 & 0 & naphthalene & phenanthrene\textsuperscript{*1} \\
        1 & 1 & 1 & 1 & naphthalene & anthracene \\
        1 & 1 & 0 & 0 & benzo[b]thiophene & naphtho[2,3-b]thiophene\textsuperscript{*2} \\
        1 & 1 & 0 & 1 & benzo[b]thiophene & naphtho[2,3-b]thiophene\textsuperscript{*2} \\
        \hline
    \end{tabular}
    \\ 
    \textsuperscript{*1} In case $n = 5$, and if the pre-determined structure is phenanthrene, this yields benzo[c]phenanthrene not forming chrysene. If naphtho[2,3-b]thiophene, phenanthro[3,2-b]thiophene is formed not forming phenanthro[2,3-b]thiophene. If naphtho[1,2-b]thiophene, phenanthro[4,3-b]thiophene is formed not forming phenanthro[1,2-b]thiophene. \\
    \textsuperscript{*2} If the pre-determined structure is naphtho[2,3-b]thiophene, naphtho[2,3-b:7,6-b']dithiophene is formed. \\
    \textsuperscript{*3} If the pre-determined structure is naphtho[2,3-b]thiophene, naphtho[2,3-b:6,7-b']dithiophene is formed.
    \label{table:N7_encoding} 
\end{table*}

\subsection{$N=7$ molecular dataset}
\label{sec:molecular_datasets}

Three main datasets containing the hole reorganization energy $\lambda$ of different heteroacene compounds were constructed and used as a testbed for the algorithms. We first constructed a dataset using a N = 7 bits encoding to verify that our framework can produce the optimal results as a proof of concept.
This encoding represents heteroacene structures containing 4 rings, which can be either benzene (no heteroatom) or thiophene (sulfur as the heteroatom). For this dataset, DFT calculated $\lambda$ values ranged from 0.063 eV to 0.345 eV. This dataset contains 54 distinct chemical structures represented by 128 bitstrings (an average of 2.4 duplicate bitstrings per chemical structure). We use 4 bits to encode which heteroatom is present in each ring, this is, we assign a value 1 if the ring is a benzene ring and a value 0 if it is a thiophene ring. The remaining 3 bits represent how the adjacent rings are fused with each other; since the chemical structures consist of an open chain of 4 rings, one bit can be assigned to describe how two rings are fused. Note that by assigning a single bit to each fusing only two connectivity options are allowed, which limits the molecular structures that are represented by the encoding. Thus, if there exist more than two ways to fuse adjacent rings, the heteroacene structures arising from such fusing need to be ignored. There might also be cases in which there is only one way of fusing two adjacent rings; in that case, both bit values will encode the same type of fusing, resulting in duplicated bitstrings (different bitstrings that correspond to the same underlying molecular structure).

We now describe in detail how the encoding was performed. As described in Figure \ref{fig:7_bits_encoding}, for the case of $N = 7$, bits $x_0$, $x_2$, $x_4$, and $x_6$ correspond to the structure of the rings, with 0 representing a thiophene ring and 1 representing a benzene ring. $x_n$ ($n = 1, 3, \text{and } 5$) shows how the consecutive rings of $x_{n-1}$ and $x_{n+1}$ are fused. 
In the case where $x_{n-1} = 0$ and $x_{n+1} = 0$ (meaning adding a thiophene ring to a thiophene ring), $x_n = 0$ leads to the structure of thieno[2,3-b]thiophene, and $x_n = 1$ leads to the structure of thieno[3,2-b]thiophene. 
In the case where $x_{n-1} = 0$ and $x_{n+1} = 1$ (meaning adding a benzene ring to a thiophene ring), both $x_n = 0$ and $x_n = 1$ lead to the same structure of benzo[b]thiophene, resulting in the presence of duplicates. 
For the case of adding a ring to a benzene ring ($x_{n-1} = 1$ and $x_{n+1} = 0, 1$), the definition of $x_n$ becomes more complicated, and the resulting fused structures are detailed in Table \ref{table:N7_encoding}.

The $N=7$ dataset is complete because it contains the $\lambda$ of all compounds represented by the $2^7$ binary strings in the dataset. 
This allows us to know beforehand which of the encoded compounds has the lowest $\lambda$, and check whether the algorithms are able to predict it. 
However, for the larger datasets using $N=16$ and $N=22$, only the $\lambda$ of a small portion of the encoded compounds is available, and therefore we cannot know which have the lowest $\lambda$. 
As described in Section \ref{sec:sequential_learning}, a sequential learning process is carried to augment these datasets in an informed way.

\subsection{$N=16$ molecular datasets}

\begin{table*}[ht]
    \centering
    \caption{Definition of fused ring structure when a ring in $x_{15}$ ($x_{15} = 0,1$) is added to a benzene ring in $x_{12}$ ($x_{12} = 1$) for $N=16$.}
    \label{table:N16_encoding}
    \begin{tabular}{cccccc}
        \hline
        $x_{10}$ & $x_{12}$ & $x_{15}$ & $x_{13}$ & $x_{14}$ &  \\
        \hline
        0 & 1 & 1 & 0 & 0 & naphtho[2,3-b]thiophene \\
        0 & 1 & 1 & 1 & 0 & naphtho[1,2-b]thiophene \\
        0 & 1 & 1 & 0 & 1 & naphtho[2,1-b]thiophene \\
        0 & 1 & 1 & 1 & 1 & naphtho[1,2-b]thiophene \\
        0 & 1 & 0 & 0 & 0 & benzo[1,2-b:5,4-b']dithiophene \\
        0 & 1 & 0 & 1 & 0 & benzo[1,2-b:4,5-b']dithiophene \\
        0 & 1 & 0 & 0 & 1 & benzo[2,1-b:3,4-b']dithiophene \\
        0 & 1 & 0 & 1 & 1 & benzo[1,2-b:3,4-b']dithiophene \\
        1 & 1 & 1 & 0 & 0 & phenanthrene\textsuperscript{*1} \\
        1 & 1 & 1 & 1 & 0 & anthracene \\
        1 & 1 & 1 & 0 & 1 & phenanthrene\textsuperscript{*1} \\
        1 & 1 & 1 & 1 & 1 & anthracene \\
        1 & 1 & 0 & 0 & 0 & naphtho[2,3-b]thiophene\textsuperscript{*2} \\
        1 & 1 & 0 & 1 & 0 & naphtho[2,3-b]thiophene\textsuperscript{*2} \\
        1 & 1 & 0 & 0 & 1 & naphtho[2,1-b]thiophene\textsuperscript{*3} \\
        1 & 1 & 0 & 1 & 1 & naphtho[1,2-b]thiophene\textsuperscript{*4} \\
        \hline
    \end{tabular}
    \\
    \textsuperscript{*1} In case $n = 5$, and if the pre-determined structure is phenanthrene, this yields chrysene not forming benzo[c]phenanthrene. If naphtho[2,3-b]thiophene, phenanthro[2,3-b]thiophene is formed not forming phenanthro[3,2-b]thiophene. If naphtho[1,2-b]thiophene, phenanthro[1,2-b]thiophene is formed not forming phenanthro[4,3-b]thiophene. \\
    \textsuperscript{*2} If the pre-determined structure is naphtho[2,3-b]thiophene, naphtho[2,3-b:7,6-b']dithiophene is formed. \\
    \textsuperscript{*3} If the pre-determined structure is naphtho[2,3-b]thiophene, naphtho[2,1-b:6,7-b']dithiophene is formed. \\
    \textsuperscript{*4} If the pre-determined structure is naphtho[2,3-b]thiophene, naphtho[1,2-b:6,7-b']dithiophene is formed.
    \label{table:x13x14_N16_encoding}
\end{table*}

The second dataset uses a N = 16 bits encoding to represent heteroacene structures, in this case with 8 rings, which can again be either benzene or thiophene rings. This dataset contains the reorganization energies (ranging from 0.072 eV to 0.286 eV) of 99 different chemical compounds represented by 926 bitstrings (approximately 9.4 duplicate bitstrings per compound). These 926 bitstrings constitute a very small portion (1.41\%) of the total number of possible binary strings with 16 bits, which is $2^{16}$ = 65536 (the bitstring solution space). We can roughly estimate the number of different heteroacene structures represented by all $2^{16}$ bitstrings by extrapolating the number of duplicates in the training dataset, resulting in 6972 encoded heteroacene structures (the chemical solution space). In a similar fashion to the N = 7 dataset, 8 bits (one per ring) are used to encode whether a ring is a benzene or a thiophene ring, while the other 8 bits store information about how two adjacent rings are fused. Since there are only 7 fusing sites in a open 8 ring chain, one of the fusion sites is assigned 2 bits so that it can encode 4 fusing options, while the rest of the fusions are represented by only 1 bit.

More precisely, for the case of $N = 16$, bits $x_0$, $x_2$, $x_4$, $x_6$, $x_8$, $x_{10}$, $x_{12}$, and $x_{15}$ represent the structure of the ring, with 0 being a thiophene ring and 1 being a benzene ring. $x_n$ ($n = 1, 3, 5, 7, 9, \text{and } 11$) shows how the two rings of $x_{n-1}$ and $x_{n+1}$ are fused, where 1 bit is assigned to describe the scheme of ring fusing. An exception exists for $x_{13}$ and $x_{14}$, which are the 2 bits used to describe the scheme of ring fusing for the two rings of $x_{12}$ and $x_{15}$.
For the case where 1 bit is assigned to describe the scheme of ring fusing ($n = 1, 3, 5, 7, 9, 10, \text{and } 11$), the definition of $x_n$ is the same as that for $N = 7$, except for three specific cases. These exceptions occur when $n = 5, 7, 9,$ or $11$ and $x_{n-3} = x_{n-1} = x_{n+1} = 1$:

\begin{enumerate}
    \item If the pre-determined structure is phenanthrene, a chrysene structure is formed instead of benzo[c]phenanthrene.
    \item If the pre-determined structure is naphtho[2,3-b]thiophene, phenanthro[2,3-b]thiophene is formed instead of phenanthro[3,2-b]thiophene.
    \item If the pre-determined structure is naphtho[1,2-b]thiophene, phenanthro[1,2-b]thiophene is formed instead of phenanthro[4,3-b]thiophene.
\end{enumerate}
These modifications were made to favor less bent structures and minimize steric hindrance within the molecule.

Next, we describe how $x_{13}$ and $x_{14}$ were defined. In the case where $x_{12} = 0$ and $x_{15} = 0$ (meaning a thiophene ring is added to a thiophene ring), $x_{13} = x_{14} = 0$ and $x_{13} = x_{14} = 1$ lead to the structure of thieno[2,3-b]thiophene, while other cases of $x_{13}$ and $x_{14}$ result in the structure of thieno[3,2-b]thiophene.
If $x_{12} = 0$ and $x_{15} = 1$ (meaning a benzene ring is added to a thiophene ring), any combination of $x_{13}$ and $x_{14}$ results in the same structure of benzo[b]thiophene, leading to duplicates.
For the case of adding a ring to a benzene ring ($x_{12} = 1$ and $x_{15} = 0, 1$), the definitions of $x_{13}$ and $x_{14}$ become more complex, and the resulting fused structures are detailed in Table \ref{table:x13x14_N16_encoding}.

\subsection{N=22 molecular dataset}

The final dataset is an expansion of the $N = 16$ bits dataset, but using $N = 22$ bits to encode the structures. It also contains heteroacene structures with 8 rings that can be either benzene or thiophene. The 6 added extra bits in the encoding are assigned to encode information about the fusing between subsequent molecules, allowing for a greater number of possible 8-ringed heteroacene structures to be represented. In this case, every scheme of ring fusion is assigned 2 bits (in total, 14 bits for encoding ring fusions) so that 4 different fusion schemes can be encoded between adjacent rings. 
Same as before, the remaining 8 bits encode the type of molecule in each ring. More precisely, $x_0$, $x_3$, $x_6$, $x_9$, $x_{12}$, $x_{15}$, $x_{18}$, and $x_{21}$ represent the structure of the ring, with 0 being a thiophene ring and 1 being a benzene ring. $x_n$ and $x_{n+1}$ ($n = 1, 4, 7, 10, 13, 16, 19$) describe how the two rings of $x_{n-1}$ and $x_{n+2}$ are fused, with their definition being the same as that for indices $x_{13}$ and $x_{14}$ for $N = 16$. 

All 99 compounds present in the $N = 16$ training dataset are also part of the $N = 22$ training dataset, and in addition to these compounds, 20 extra compounds were further added to explore the new subspace spanned by the 6 extra bits. In total, this dataset contains the $\lambda$ of 119 different chemical compounds ($\lambda$ ranging from 0.072 eV to 0.286 eV) represented by 24,512 bitstrings (0.58\% of the bitstrings solution space), having 206 duplicate bitstrings per compound on average. The bitstring solution space is now approximately 4.2 million ($2^{22}$), which represents a total of approximately 20,361 compounds, calculated again by extrapolating the number of duplicate bitstrings in the training dataset.

\subsection{Data processing}
\label{sec:data_preprocessing}

To train the models in the learning phase of the algorithms, we split the data into a training set and a test set in order to measure the generalization capabilities of the models to unseen data. 
Recall that our input data are binary strings $x$ with an assigned scalar value each (the hole reorganization energy $\lambda$). 
Each binary string represents a single chemical compound, but the opposite is not true as each chemical compound may be represented by more than one bitstring (duplicates).

Due to the presence of duplicate bitstrings, one needs to be careful when splitting the data into a training and test set. 
In fact, we construct two separate test datasets. 
The first is the \textit{non-isolated test dataset}, which is constructed with duplicates of bitstrings that are present in the training dataset, meaning that the underlying chemical compound is actually seen during training. 
Therefore, the performance on this non-isolated data only measures the ability of the model to identify the symmetries in the encoding giving rise to the duplicates, but it does not truly measure generalization to unseen chemical compounds. 
The second dataset is the \textit{isolated test dataset}. This dataset contains \textit{all} duplicates of a number of randomly picked chemical structures that are not part of the training dataset, meaning that no duplicates of those structures are seen during training. 
Therefore, the performance on this dataset truly measures the ability to generalize to unseen chemical compounds. 

For all datasets we do a 8:1:1 split (train/isolated test/non-isolated test) of the data. Furthermore, the data is always normalized to the $(0,1)$ interval before it is provided as input to the models.

\section{Algorithmic framework}
\label{sec:algorithmic_framework}

\subsection{Extremal learning}
The algorithmic framework used in this work is known as \textit{extremal learning} \cite{Varsamopoulos2024,patel2021extremallearningextremizingoutput}. Extremal learning algorithms aim to find the input $x_{opt}$ to a hidden function $f: x\in X \rightarrow y\in Y$ which extremizes the output of the function $y$, only given partial input-output training data $\{(x_i,y_i)\}$. 
In our case, both the input and the output of the function are one-dimensional, but extremal learning can be straightforwardly generalized to higher dimensions. 
Usually, this is achieved via two distinct phases, a \textit{learning} (or modeling) phase and a \textit{selection} (or extremization) phase.
In the learning phase, a (quantum) machine learning model is trained on the partial training data to build an approximate model $\hat{f}(x)$ of the hidden function $f(x)$.
Once $\hat{f}(x)$ is learned, an extremization procedure is performed in the selection phase to find $\hat{x}_{opt}$ that extremizes the output of the approximate model $\hat{f}(x)$. 
If $\hat{f}(x)$ is sufficiently accurate and the extremization process succeeds, $\hat{x}_{opt}$ will correspond (or be close) to the optimal input of the hidden function, $\hat{x}_{opt}= x_{opt}$.

Extremal learning algorithms are naturally suited for the task of material discovery where the fitness of a molecule can be faithfully described by one or more figures of merit. In this case, the inputs are the considered family of chemical structures, represented via the chosen encoding, while the output can be any figure of merit that is desirable to extremize; in our case, we wish to minimize the hole reorganization energy $\lambda$. 
Different extremal learning algorithms can employ different strategies for the learning and selection phase.  
Depending on the particular algorithm, both phases can be performed via classical or quantum computation. 

Here we employ two quantum-enhanced extremal learning algorithms, namely \textit{Quantum Extremal Learning} (QEL) \cite{Varsamopoulos2024} and \textit{Factorization Machine + Quantum Approximate Optimization Algorithm} (FM+QAOA) \cite{Kitai2020}.
We use the term "quantum-enhanced" to remark the fact that these algorithms employ both quantum and classical computation at some point in their workflow.
QEL employs a QML algorithm in the learning phase and a classical method in the selection phase, while FM+QAOA follows an opposite approach using a classical ML model in the learning phase and a QML algorithm in the selection phase.
Next, we describe both algorithms in detail.

\subsection{FM+QAOA}

The first extremal learning algorithm is named \textit{Factorization Machine}+ \textit{Quantum Approximate Optimization Algorithm}, or FM+QAOA for short. It employs a classical ML model (Factorization Machine) in the learning phase and a quantum algorithm (QAOA) in the selection phase. 
This method was first proposed in \cite{Kitai2020}, where it was successfully applied to a material design problem. 

\subsubsection{Learning phase: Factorization Machines}

A Factorization Machine (FM) is a supervised classical machine learning algorithm \cite{Koren2009} commonly used in recommendation systems \cite{Gomez2016}.
Although in principle a FM model can have any arbitrary order $d\geq 2$, which allows to capture all $d$-th order correlations between the input features $x_i$, here we restrict the FM to be second order ($d=2$). 
Higher order FMs \cite{blondel2016higherorderfactorizationmachines} may provide a more accurate model if the data presents significant higher order correlations, but in the context of the FM+QAOA algorithm it would significantly increase the complexity of the quantum circuit implementing QAOA.
Moreover, a second order approximation is enough in many cases, particularly in extremal learning algorithms where high accuracy is not required for the selection phase to succeed.

For binary input vectors $x\in \{0,1\}^N$ a second order FM can be mathematically written as 

\begin{equation}
    \hat{f}_{\text{FM}}(x) = w_0 + \sum_{i=1}^{N} w_i x_i + \frac12\sum_{i,j=1}^{N}w_{ij} x_i x_j ,
    \label{eq:fm_order_2}
\end{equation}

where $w_{ij} = (\boldsymbol{v}\cdot \boldsymbol{v})_{ij} = \sum_{f=1}^{K} v_{if} v_{jf}$ are the elements of the factorized matrix of quadratic coefficients. 
A second order FM captures all single and pairwise interactions between the input features $x_i$. 
The pairwise interactions are modeled by factorizing the coefficients $w_{ij}$ in a embedding matrix $\boldsymbol{v}$ of size $(N,K)$, where $K$ is a hyperparameter controlling the degree of factorization. 
Reducing $K$ can help control overfitting as the number of parameters is also reduced, although in this work we fix $K=N$, thus not performing any factorization.

\subsubsection{Selection phase: QAOA}

The goal of the selection phase is to find the optimal solution $x_{opt}$ that extremizes the hidden function $f(x)$, using the trained FM model $\hat{f}_{FM}(x)$ as a proxy.
Without loss of generality, we can assume that the solution $x_{opt}$ is the minimum of $f(x)$ (if $x_{opt}$ is such that it maximizes $f(x)$ then the problem can always be reformulated as $f(x)\rightarrow -f(x)$).
The task of finding the optimal input $x_{opt}$ minimizing Eq. (\ref{eq:fm_order_2}) can be formulated as a Quadratic Unconstrained Binary Optimization (QUBO) problem \cite{blekos2024, Lucas_2014}.
Any QUBO problem can be mapped to an Ising Hamiltonian via the transformation $x_i\rightarrow \sigma_z$, where $\sigma_z$ is the Z-Pauli matrix.
The resulting Ising Hamiltonian,

\begin{equation}
    H_{\text{Ising}} = w_0 + \sum_{i=1}^{N} w_i Z_i + \frac12 \sum_{i,j=1}^{N} w_{ij} Z_i Z_j ,
    \label{eq:ising_hamiltonian}
\end{equation}

encodes the solution of the QUBO problem in its ground state. 

Therefore, the task of finding $x_{opt}$ can be translated to finding the ground state of $H_{\text{Ising}}$.
A wide range of quantum algorithms for ground state preparation have been proposed, both for fault tolerant \cite{Lin_2020} and Noisy Intermediate-Scale Quantum (NISQ) \cite{Tilly_2022} devices.
Here we use the Quantum Approximate Optimization Algorithm (QAOA) \cite{farhi2014, Zhou_2020, blekos2024}, a variational quantum algorithm motivated by a trotterized version of the quantum adiabatic algorithm.
The algorithm employs two different operators, a \textit{cost operator} $U_C(\gamma)$ and a \textit{mixing operator} $U_{M}(\beta)$ defined as

\begin{equation}
\begin{split}
U_C(\boldsymbol{\gamma}) &= e^{i\boldsymbol{\gamma} H_C} \\
U_M(\boldsymbol{\beta}) &= e^{i\boldsymbol{\beta} H_M}
\end{split}
\end{equation}

where $H_C$ and $H_M$ are the cost and mixing Hamiltonians, respectively, and $(\boldsymbol{\gamma}, \boldsymbol{\beta})$ are variational parameters.
The cost Hamiltonian encodes the solution in its ground state; in our case, it is the Ising Hamiltonian defined by the Factorization Machine, $H_C = H_{\rm Ising}$. 
For the mixing Hamiltonian $H_M$ one may select any Hamiltonian which does not commute with the cost Hamiltonian i.e. $[H_C, H_M]\neq 0$. In our case, we choose the standard $H_M = \otimes_i \sigma_{x,i}$, where $\sigma_{x,i}$ is the X-Pauli matrix applied on qubit $i$.

The final state is prepared by applying $L$ successive layers of the cost operators and mixing operators applied to the initial state, which is the ground state of the mixing Hamiltonian, in our case $\ket{+}$:

\begin{equation}
\ket{\psi(\boldsymbol{\gamma}, \boldsymbol{\beta})}= U_C(\gamma_L)U_M(\beta_L) \ldots U_C(\gamma_1)U_M(\beta_1) \ket{+}.
\end{equation}

The variational parameters $\boldsymbol{\gamma}, \boldsymbol{\beta}$ are trained with a classical optimizer to minimize the expectation value of the cost Hamiltonian $\bra{\psi(\boldsymbol{\gamma}, \boldsymbol{\beta})} H_C \ket{\psi(\boldsymbol{\gamma}, \boldsymbol{\beta})}$ via a quantum-classical feedback loop.
If the algorithm converges, sampling the final state in the computational basis should give the state corresponding to the optimal solution $\ket{x_{opt}}$ with the highest probability.
In our simulations we simulate the final sampling process with finite shot noise.

Many variants of QAOA have been proposed to improve its performance and trainability; however, in this work, we employ the originally proposed version explained above.
The only algorithmic trick we use relates to the initialization of the variational parameters $(\boldsymbol{\gamma},\boldsymbol{\beta})$. 
In particular, to train a QAOA circuit of depth $L$, we progressively extend the depth of the circuit from a single layer to $L$ layers, using the optimal values found with depth $l$ as the initial values for the first $l$ layers of the circuit of length $l+1$. 
This means that in order to train a QAOA circuit of depth $L$, we first need to train $L-1$ circuits with depths $l=1\ldots L-1$, which forces a significant computation overhead. 
However, this initialization strategy greatly benefits convergence by avoiding local minima \cite{Lee_2021}.
We note that other initialization strategies with lower overhead and similar performance might be available.
We leave other possible extensions of QAOA as future continuation of this work.

\subsection{Quantum Extremal Learning}

The second extremal learning algorithm is the \textit{Quantum Extremal Learning} algorithm proposed in \cite{Varsamopoulos2024}. 
In the original paper, both the learning and selection phases employ a quantum approach leveraging Quantum Neural Networks (QNNs) for both continuous and discrete data. 

However, when the input data are discrete the selection method proposed in \cite{Varsamopoulos2024} converges to exactly a single solution. To fuel the sequential learning process to expand the $N=16$ and $N=22$ bits datasets  (explained in Section \ref{sec:sequential_learning}), we require the algorithm to generate 50 candidate solutions in each iteration.
For this reason, we only use the original selection method for the smallest N=7 dataset, where no sequential learning process is carried since the dataset is already complete.
For the larger datasets, we instead use a brute-force method for selecting the best candidates, explained in Section \ref{sec:qel_selection_phase}. 
Note that neither the brute-force method used here nor the selection method used in \cite{Varsamopoulos2024} are efficient; further work is required to design a more efficient selection procedure in QEL.

\subsubsection{Learning phase: Quantum Neural Networks}

The learning phase in QEL is carried by a Quantum Neural Network (QNN), which we denote as the \textit{learning QNN}. 
We define a QNN as a variational quantum circuit acting on a $N$-qubit register. The quantum circuit used here consists of two parts: a feature map $U_{\mathbf{x}}$ and a variational ansatz $U_{\mathbf{\theta}}$.
The feature map $U_{\mathbf{x}}$ is used to encode the classical input data into the circuit:

\begin{equation}
    \ket{x} = U_{\mathbf{x}} \ket{0}^{\otimes N} ,
\end{equation}

where $\ket{x}$ is the quantum state resulting from the action of the feature map on the initial state $\ket{0}^{\otimes N}$.
Since in our case the input data are binary strings of length $N$, we use a basis encoding where a bitstring $x$ is mapped onto the corresponding computational basis state $|x\>$ (e.g. the bitstring $x=0011$ is mapped into the the computational state $x=|0011\>$).
This feature map can be easily implemented via a layer of $\sigma_x$ gates acting on the qubits that correspond to a bit value 1, leveraging the fact that $\ket{1} =\sigma_x \ket{0}$. 

After encoding the classical data into $\ket{x}$, a variational ansatz $U_{\boldsymbol{\theta}}$ is applied. In our case, $U_{\boldsymbol{\theta}}$ is a Hardware-Efficient Ansatz (HEA) \cite{Kandala_2017} consisting of layers of single qubit rotations $R_{z}(\theta_{i,l,1})R_{x}(\theta_{i,l,2})R_{z}(\theta_{i,l,3})$ acting on qubit $i$ and layer $l$, interleaved with entangling blocks of chained CNOTs. 
Information is extracted from the circuit by performing a measurement $\hat{O}$ at the end, in our case the average local magnetization $\hat{O} = \frac1N \sum_i \sigma_z^{(i)}$.
The variational parameters $\boldsymbol{\theta}$ are optimized such that the expectation value of the measurement operator matches the hidden function $f(x)$ for each bitstring $x$

\begin{equation}
\label{eq:expectation_value}
    f(x) \approx \hat{f}(x) =  \bra{x}U_{\boldsymbol{\theta}}^\dagger\hat{O}U_{\boldsymbol{\theta}}\ket{x},
\end{equation}

The predicted function $\hat{f}(x)$ can be differentiated with respect to $\boldsymbol{\theta}$ through the parameter shift rule \cite{Mitarai_2018} (although in our numerical experiments we use automatic differentiation as we emulate the quantum algorithms in classical hardware), so standard gradient-based optimization routines can be used to find the optimal values of $\boldsymbol{\theta}$.
We use additional trainable parameters for the scaling and shifting of the output, $\theta_{\rm{scale}}\cdot \hat{f}(x)+\theta_{\rm{shift}}$, and we always normalize the data to the be in the $[0,1]$ range before feeding it to the model.

An important benefit of using a QNN compared to a second order FM in the learning phase is that a QNN can model correlations of arbitrary order between the input features $x_i$, given an appropriate ansatz. 
Although a Factorization Machine can also be generalized to arbitrary order $d$ (mapping the data to a general $d$-order $\sigma_z$ Hamiltonian instead of an Ising Hamiltonian), this may require very deep circuits when implementing the cost operator $U_c = e^{i\gamma H_c}$ in the QAOA circuit, rendering the latter approach very computationally demanding for large $d$.

\subsubsection{Selection phase}
\label{sec:qel_selection_phase}

Different selection methods are required in QEL depending on whether the input data is continuous or discrete. 
The continuous case is straightforward, as we can leverage the fact that the right-hand side of Eq. (\ref{eq:expectation_value}) is differentiable with respect to the feature map parameters $\mathbf{x}$ and run a gradient-descent optimization routine to find the optimal input $\mathbf{x}_{\rm{opt}}$.
However, when the data is discrete, the differentiability vanishes, so an alternative approach needs to be employed to find the extremal input.

Throughout this work, we use two different selection methods tailored to discrete data, depending on the dataset we are dealing with. 
For the smallest dataset $N=7$, we follow the approach presented in \cite{Varsamopoulos2024}, where the discrete feature map in the quantum circuit is substituted by a \textit{selection QNN}, whose action is denoted by the unitary  
$U_{\boldsymbol{\mathcal{X}}}$, with \textit{continuous} variational parameters $\boldsymbol{\mathcal{X}}$.
This unitary generates the state

\begin{equation}
    |\mathcal{F}(\boldsymbol{\mathcal{X}})\>=U_{\boldsymbol{\mathcal{X}}}(\boldsymbol{\mathcal{X}})\ket{0}
\end{equation}

After convergence, this state should have major support on the quantum state corresponding to the optimal input $x_{\rm{opt}}$, i.e. $\ket{x_{\rm{opt}}} \approx |\mathcal{F}(\boldsymbol{\mathcal{X}}_{\rm{opt}})\>$.
In our case, we choose $U_{\boldsymbol{\mathcal{X}}}$ to be a Hardware-Efficient Ansatz with the same architecture as in the learning phase.
Since $U_{\boldsymbol{\mathcal{X}}}$ is differentiable with respect to $\boldsymbol{\mathcal{X}}$, a gradient-descent optimization routine can be employed to find the optimal parameters $\boldsymbol{\mathcal{X}}_{\rm{opt}}$. 
Once it has successfully converged, the optimal parameters can be extracted by sampling the state $|\mathcal{F}(\boldsymbol{\mathcal{X}}_{\rm{opt}})\>$.

However, in order for $U_{\boldsymbol{\mathcal{X}}}$ to represent the optimal solution, the optimization routine needs to be adjusted to account for the fact that the data is discrete. 
To showcase this adjustment, let us write the state $|\mathcal{F}(\boldsymbol{\mathcal{X}})\>$ as a linear combination of computational basis states:

\begin{equation}
    |\mathcal{F}(\boldsymbol{\mathcal{X}})\> = \sum_{i=1}^{2^N} c_i (\boldsymbol{\mathcal{X}}) |i\>.
    \label{eq:extremizer_state}
\end{equation}

where $c_i(\boldsymbol{\mathcal{X}}) \in \mathbb{C}$ and $\{|i\>\}_{i=1}^{2^N}$ are the computational basis states of $N$ qubits. The loss we aim to minimize can then be written as

\begin{equation}
\begin{aligned}
        \mathcal{L}_\text{ext} (\boldsymbol{\mathcal{X}}) &= \<\mathcal{F}(\boldsymbol{\mathcal{X}})| U_{\pmb\theta}^{\dag}\hat{O} U_{\pmb\theta}|\mathcal{F}(\boldsymbol{\mathcal{X}})\> \\
        &= \sum_{i=1}^{2^N} |c_i(\boldsymbol{\mathcal{X}})|^2 \<i|  U_{\pmb\theta}^{\dag}\hat{O} U_{\pmb\theta} |i\> \\
        & + 2\sum_{i>j}^{2^N} \text{Re}(c_i^*(\boldsymbol{\mathcal{X}}) c_j(\boldsymbol{\mathcal{X}})) \<i|  U_{\pmb\theta}^{\dag}\hat{O} U_{\pmb\theta} |j\> \\
        &= T_{\text{diag}} + T_{\text{cross}}
\end{aligned}
\end{equation}

The loss can be written as the sum of two terms: a \textit{diagonal} term $T_{\text{diag}}$ containing the diagonal matrix elements $\<i| U_{\pmb\theta}^{\dag}\hat{O} U_{\pmb\theta} |i\>$ and a \textit{cross} term $T_{\text{cross}}$ containing the non-diagonal elements $\<i| U_{\pmb\theta}^{\dag}\hat{O} U_{\pmb\theta} |j\>$. 
Only minimizing the diagonal term contributes to the state $|\mathcal{F}(\boldsymbol{\mathcal{X}})\>$ becoming the optimal state. 
Therefore, we only use the diagonal term, i.e. we minimize $\mathcal{L}_\text{ext} (\boldsymbol{\mathcal{X}})= T_{\text{diag}}(\boldsymbol{\mathcal{X}})$. 
Computing only the diagonal term involves calculating the exponential sum $\sum_{i=1}^{2^N} |c_i(\boldsymbol{\mathcal{X}})|^2 \<i| U_{\pmb\theta}^{\dag}\hat{O} U_{\pmb\theta} |i\> $ in each optimization step, and therefore it is not an efficient procedure.

A consequence of the above adjusted optimization process is that the state $|\mathcal{F}(\boldsymbol{\mathcal{X}})\>$ usually converges to only the single best computational state.
In general, it is ideal that the final state has no support on other states that are close to optimal, but for the purposes of the sequential learning process described in Section \ref{sec:sequential_learning}, we would like the selection algorithm to output the best $k$ states (we fix $k=50$ in our case) so that we can expand the dataset with a meaningful number of candidates at each iteration. 
In order to circumvent this issue, for the large datasets with $N=16$ and $N=22$, we employ a simpler selection routine based on a brute-force approach sorting the bitstrings by their corresponding predicted energy, based on the trained quantum model.
Although this naive approach is also not efficient, and thus it cannot be scaled to a large number of qubits, it allows us to select an arbitrary number of candidates, so that the sequential learning process can be carried.
Furthermore, using the brute-force approach, we are guaranteed to find the correct extremal inputs of the model resulting from the learning phase.

\subsection{Sequential learning process}
\label{sec:sequential_learning}

Sequential learning (SL) \cite{ rohr2020benchmarking, Palizhati2022, volker2021sequential, Balachandran2016} refers to the process of iteratively updating a model to guide (physical or computational) experiments.
In a SL process, the model actively solicits new data to improve its performance, creating a feedback loop between the experimental results and the model's training process. 
SL has been shown to be useful in materials science as it can reduce the number of compounds that need to be evaluated through DFT calculations \cite{D0SC01101K, volker2021sequential} or chemical synthesis \cite{Noack2021GaussianPF}. 
Since these evaluations usually constitute the most expensive part of the material discovery pipeline, SL can greatly speed up material discovery algorithms.

In our case, SL offers an ideal avenue to expand the limited amount of initial training data available for the $N=16$ and $N=22$ datasets.
Recall that the amount of data initially available to us is just 927 bitstrings (corresponding to 99 chemical structures) for the $N=16$ dataset and 24512 bitstrings (119 chemical structures) for the $N=22$ dataset, representing 1.41\% and 0.58\% of the bitstring solution space, respectively.
To circumvent this issue, we follow a SL process where we use the predictions of our algorithms to iteratively expand the training data.
At each iteration, a set of 50 candidate bitstrings is generated, corresponding to the chemical structures predicted to be most optimal, i.e., to have the lowest predicted reorganization energy $\lambda^{\rm{(pred)}}$.
Next, DFT methods are used to compute the true reorganization energy $\lambda^{\rm{(dft)}}$ of the chemical compounds represented by those candidate bitstrings. 
At this stage, all duplicate bitstrings representing the predicted chemical structures are identified and incorporated into the dataset.
By augmenting the training dataset with the compounds predicted to have the lowest $\lambda^{\rm{(pred)}}$ in the previous iteration, we aim to bias the model towards predicting low-$\lambda$ compounds.
Note that despite the fact that 50 bitstrings are generated at each iteration, some of the predicted bitstrings might be duplicates which represent the same chemical structure. Therefore, usually less than 50 new unique chemical compounds are added in each iteration. On the other hand, due to the added duplicates, more than 50 new bitstrings are added at each iteration.

\begin{figure}[htb]
    \centering
    \includegraphics[width=1.0\linewidth]{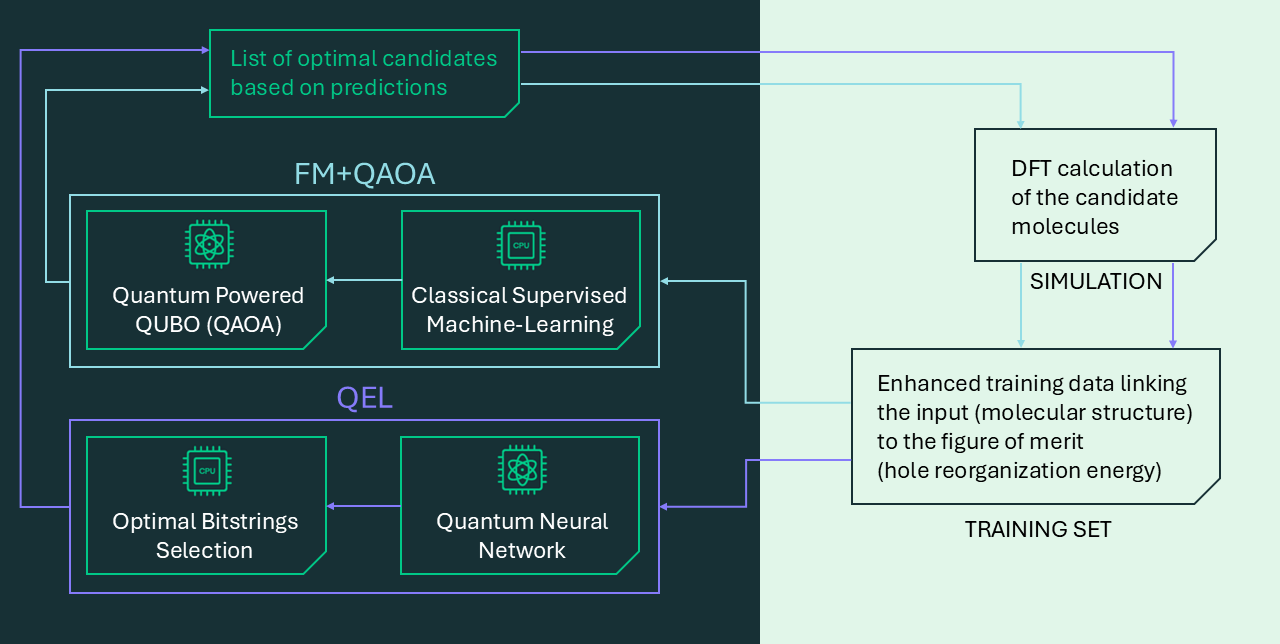}
    \caption{Schematic diagram of the sequential learning process employed for the $N=16$ dataset (where both FM+QAOA and QEL algorithms were used) and $N=22$ dataset (where only FM+QAOA was applied).
    The diagram shows the flow of data in the sequential learning process, which happens independently for the two algorithms. 
    Firstly, the model is trained with the available data in the learning phase (the Factorization Machine in the FM+QAOA and the Quantum Neural Network in QEL), and then the selection phase uses the trained model to generate the list of optimal candidates. In the FM+QAOA, this is done via QAOA, while in QEL this is done via a brute-force approach. 
    }
    \label{fig:iterative_process_diagram}
\end{figure}

As a result of the SL process, both the size and quality of the dataset increase with each iteration, as the added compounds are predicted by the algorithm to have low $\lambda$. Consequently, the algorithms should improve at predicting low-$\lambda$ structures as the SL process advances.
This SL process is done \textit{independently} for both the FM+QAOA and the QEL algorithm, meaning that no data exchange occurs between them and each algorithm has no knowledge of the predictions of the other.
This is made explicit in Figure \ref{fig:iterative_process_diagram} by the two independent arrows representing the independent flow of information for the FM+QAOA (blue arrows) and QEL (purple arrows).

At every iteration, 30 instances of each algorithm were run, each instance with different portions of the data assigned to the train and test sets as well as different initial parameters.
Out of these 30 instances, we filtered the instances achieving a mean absolute error (MAE) on the isolated test dataset (see Section \ref{sec:data_preprocessing}) greater than a certain threshold. 
Given that our only focus is to build models that generalize well to the low-$\lambda$ regime, we only focus on the MAE of compounds with a $\lambda<0.1$ eV.
The training process of each instance was also stopped at the point achieving the lowest MAE in these low-$\lambda$ compounds, in order to improve the accuracy of the models in this regime.
We use the MAE and not the $R^2$ score because the latter may not be accurate when calculated on samples drawn from a narrow range (the variance in the denominator might diverge), which is the case when we focus solely on the low-$\lambda$ regime.

\section{Numerical results}
\label{sec:results}

In this Section, we discuss the performance of the FM+QAOA and QEL algorithms applied to the three molecular datasets described in Section \ref{sec:molecular_datasets}.
The $N=7,16$ datasets were benchmarked with both the FM+QAOA and the QEL algorithms, while the $N=22$ dataset was only tested with the FM+QAOA algorithm.

The numerical results presented here were computed in a classical computer cluster employing CPU and GPU resources, and all quantum algorithms were programmed with Qadence \cite{qadence2024pasqal}. 
Derivatives were computed through automatic differentiation, enabled through Qadence's Pytorch backend \cite{Pyqtorch, paszke2017automatic}.
No noise was considered other than finite shot noise when sampling the final state in QAOA.
Regarding training, all models were trained with the ADAM optimizer \cite{kingma2017adammethodstochasticoptimization} using a decaying learning rate strategy.

\subsection{N=7 dataset}

We start with the smallest dataset built with the N=7 bits encoding. 
The dataset contains 54 different chemical structures represented by 128 bitstrings of length 7.
This is the only dataset where we have access to the figure of merit (the DFT-calculated hole reorganization energy $\lambda^{\rm (dft)}$) for all bitstrings in the solution space. This allows us to perform a thorough analysis of the algorithms' performance by leveraging the fact that we know the optimal solution beforehand.

\subsubsection{Hyperparameter tuning}

To ensure a fair comparison between the two algorithms, we conducted a hyperparameter search to identify optimal hyperparameter settings for each training dataset size.
For the FM+QAOA algorithm, the most important hyperparameter is the number of layers of the QAOA algorithm. 
Theoretically, increasing the QAOA depth should always give greater accuracy, with asymptotical guarantees at the infinite depth limit. 
However, in practice the variational parameters might not converge to optimal values because the training process might become very hard for large depths. 
For the $N=7$ dataset, our numerical experiments revealed that the performance of the QAOA algorithm was not significantly influenced by the depth of QAOA. 
In fact, for all training dataset sizes, performance plateaued when the QAOA depth exceeded 4 layers. 
Consequently, the depth was fixed at 4 layers for the remainder of the experiments with $N=7$.
Finally, we used $n_{\rm{shots}} = 10^3$ shots to sample the final state of QAOA.

The main hyperparameter in the QEL algorithm is the depth of the Quantum Neural Networks (QNNs) used in the learning and selection phase.
In this case, we performed a grid search to study the performance of the QNNs with depth ranging from 2 to 8 layers by looking at the cross-validation $R^2$ score averaged over 25 instances for each depth. 
Table  
\ref{table:N7_optimal_hyperparameters} shows the optimal number of layers for different training dataset sizes. 
We observe that as the amount of training data increases, the optimal depth of the learning QNN also increases.
This is likely because training a QNN that is too deep on insufficient data can lead to overfitting. Increasing the amount of training data allows for deeper QNNs, which are more expressive, while avoiding overfitting. 
On the other hand, the optimal depth of the selection QNN remains approximately constant for different training dataset sizes. 
This is likely because the selection QNN does not aim to model any data, but rather to represent the quantum state corresponding to the optimal solution. 
Since we are only minimizing a function instead of learning the function, there is no need to increase the depth when the amount of training data increases.
For the rest of this Section we fix the depths of the QNNs to the values shown in Table \ref{table:N7_optimal_hyperparameters}.

\begin{table}[]
\centering
\begin{tabular}{|c|c|c|c|c|c|}
\hline
$n_{\text{train}}/n_{\text{data}}$ & \ 0.2 \  & \ 0.4 \  & \ 0.6 \  & \ 0.8 \  & \ 1.0 \  \\ \hline
Optimal depth learning QNN         & 4   & 5   & 6   & 7   & 7   \\ \hline
Optimal depth selection QNN        & 3   & 4   & 3   & 3   & 4   \\ \hline
\end{tabular}
\caption{Optimal depth for the QNNs used in the learning and selection phases of QEL for the $N=7$ dataset.}
\label{table:N7_optimal_hyperparameters}
\end{table}

\begin{figure}[htb]
    \centering
    \includegraphics[width=0.84\linewidth]{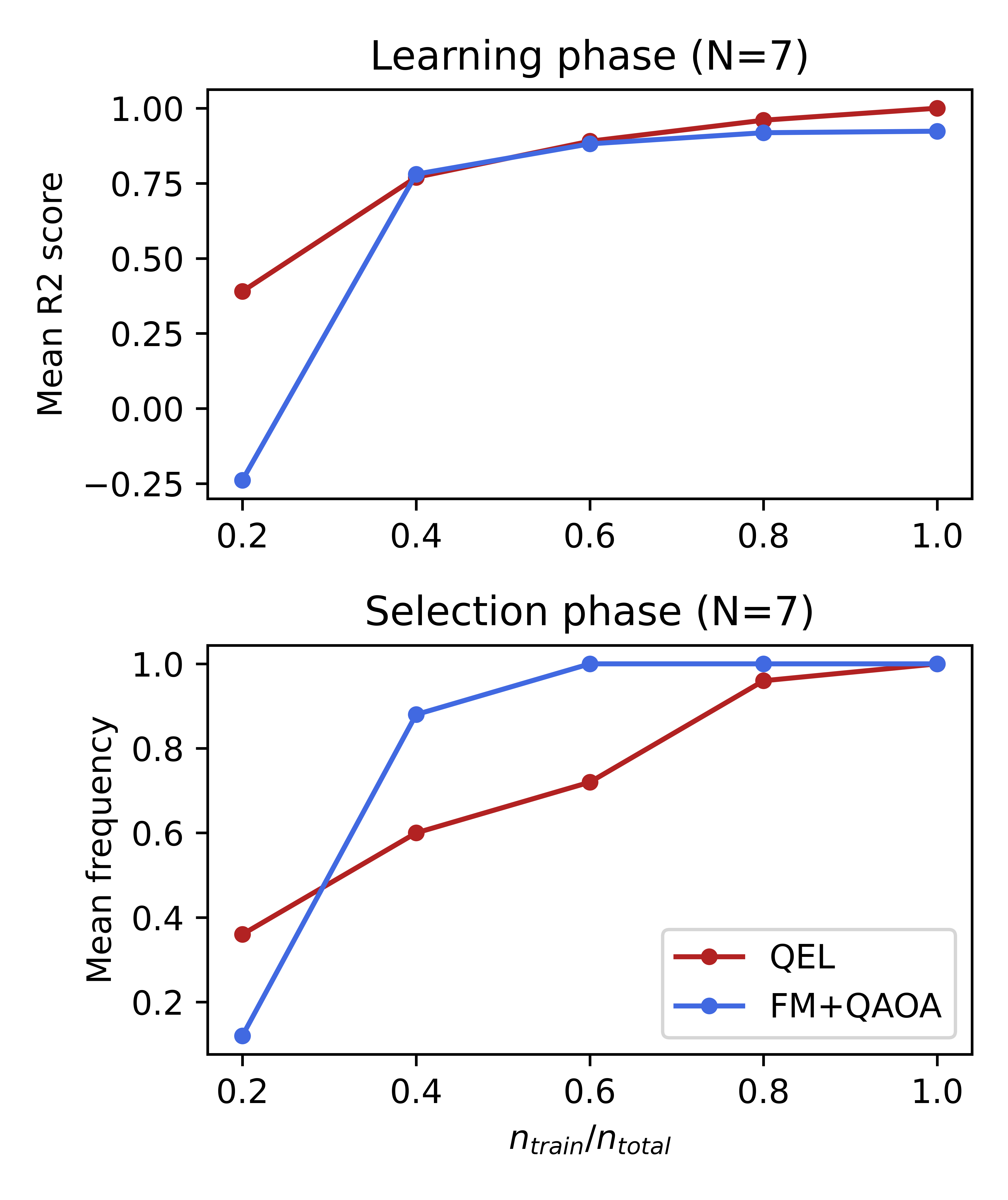}
    \caption{Performance of the FM+QAOA and QEL algorithms when trained on the N=7 bits dataset with different training dataset sizes $n_{train}/n_{total}$. The results are averaged over 25 instances. (top) Mean $R^2$ score of the models in the learning phase calculated on the complete dataset. (bottom) Mean normalized frequency of finding the optimal solution out of all seeds.} 
    \label{fig:methods_comparison_molecular_N7}
\end{figure}

\subsubsection{Results}

Figure \ref{fig:methods_comparison_molecular_N7} shows the performance of both algorithms with the optimal hyperparameter configuration.
We measure the performance for different ratios $n_{\text{train}}/n_{\text{total}}$, where $n_{\text{train}}$ is the number of training samples and $n_{\text{total}}$ is the total number of samples (in this case $2^7=128$).

For each ratio, 25 instances of the algorithms were trained with different portions of the data assigned to the train/test sets and different initial parameters. 
The top plot of Figure \ref{fig:methods_comparison_molecular_N7} showcases the performance of the models in the learning phase using as a figure of merit the average $R^2$ score of the 25 seeds. The $R^2$ score is defined as 

\begin{equation}
    R^2 =1 - \frac{\sum_x{\left(\lambda^{(\rm{dft})}(x)-\lambda^{\text{(pred)}}(x)\right)^2}}{\sum_x{\left(\lambda^{\rm{(dft)}}(x)-\langle \lambda^{\rm{(dft)}}\rangle\right)^2}},
    \label{eq:R2_score}
\end{equation}

where $x$ are the input bitstrings in the dataset, $\lambda^{\rm{(pred)}}$ is the predicted reorganization energy by the model and $\lambda^{\rm{(dft)}}$ is the DFT-calculated reorganization energy.
The mean $R^2$ score achieved by QEL is better than the FM when using only 20\% of the data for training, which indicates better generalization. However, the QEL algorithm only outperforms the FM algorithm for very small or very large training datasets, and achieves a comparable performance for medium sizes $n_{train} / n_{data}$ = 0.4 and 0.6. 
An important observation is that the FM is not able to achieve a perfect score even when using all data for training ($R^2 = 0.92$), because the FM cannot model the higher order correlations present in the dataset. On the other hand, the QNN in the QEL algorithm does achieve $R^2 = 1.00$ when trained with the whole dataset because it is able to model the higher order correlations successfully.

The bottom plot in Figure  \ref{fig:methods_comparison_molecular_N7} shows the results in the selection phase, using as a figure of merit the normalized frequency with which the algorithms found the optimal solution. This is, out of the 25 instances, we count how many times the algorithm predicted the compound with the lowest reorganization energy as the most optimal, and then normalize the result. 
In this case, the FM+QAOA algorithm outperforms the QEL algorithm for all of the training dataset sizes except for the lowest amount of training data, $n_{\text{train}}/n_{\text{total}}=0.2$. 
This also indicates that the QEL algorithm can generalize better when very few training data is used, but on the other hand it performs worse with larger amounts of training data.

An important take-away of Figure \ref{fig:methods_comparison_molecular_N7} is that the models in the learning phase are not required to be very accurate for an extremal learning algorithm to succeed in finding close-to-optimal solutions.
In fact, even though QEL was able to model the molecular dataset better than the FM model, the extremization results of QEL are worse than with FM+QAOA for most ratios except when $n_{\text{train}}/n_{\text{total}}=0.2$. 
The reason behind is that as long as the model correctly captures the position of the global extremum point, that should be enough for the selection method to find the solution if the selection process succeeds.

\subsection{N=16 dataset}

\begin{figure}[tb]
    \centering
    \includegraphics[width=1.0\linewidth]{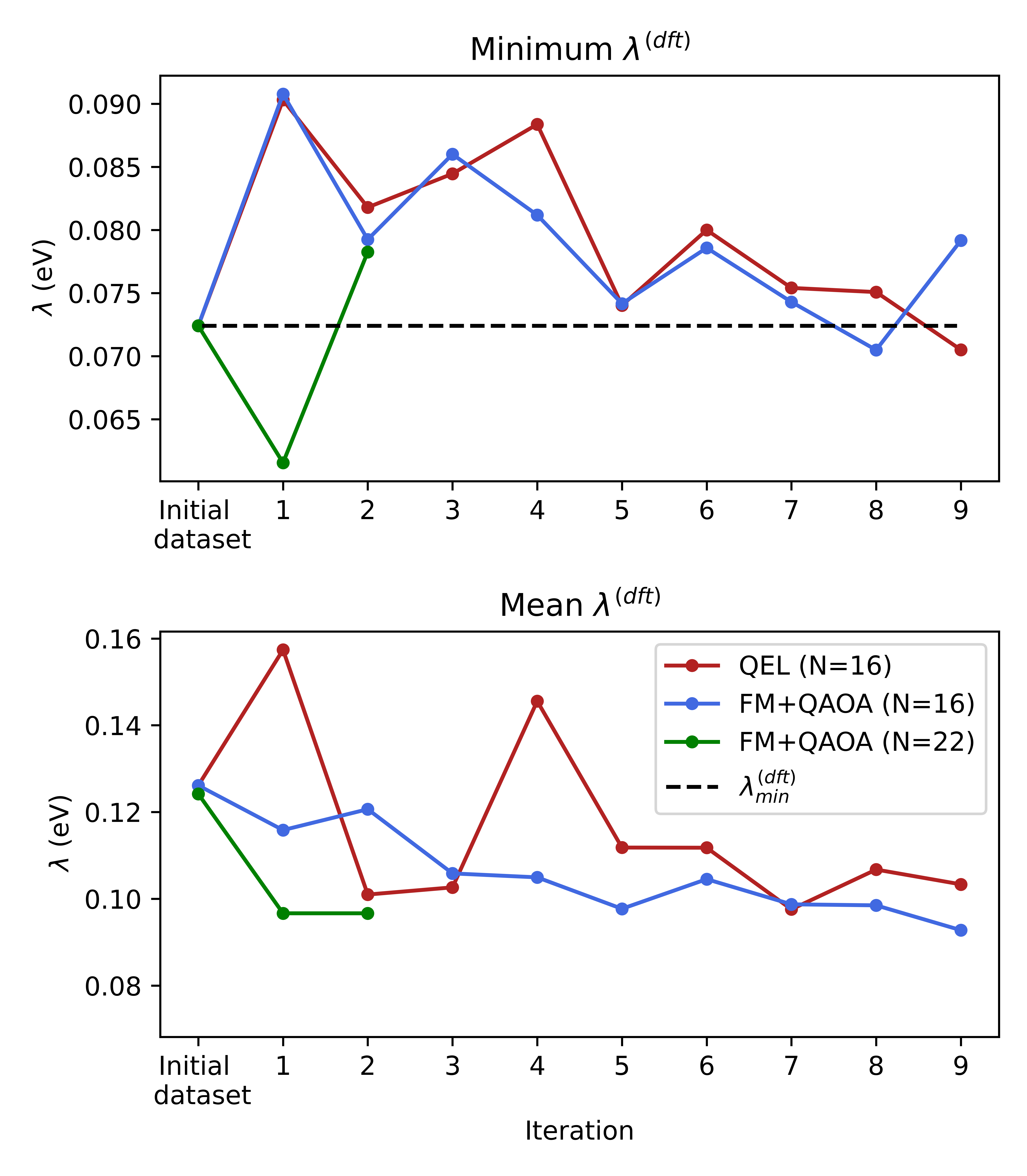}
    \caption{Minimum (top) and mean (bottom) DFT-calculated hole reorganization energy $\lambda^{\rm (dft)}$ of the predicted candidates in each iteration for QEL and FM+QAOA. The minimum $\lambda$ present in the initial dataset is denoted as $\lambda_{\rm{min}}^{\rm{(dft)}}$}
    \label{fig:means_and_mins}
\end{figure}

\begin{figure*}[htb]
    \centering
    \includegraphics[width=0.80\linewidth]{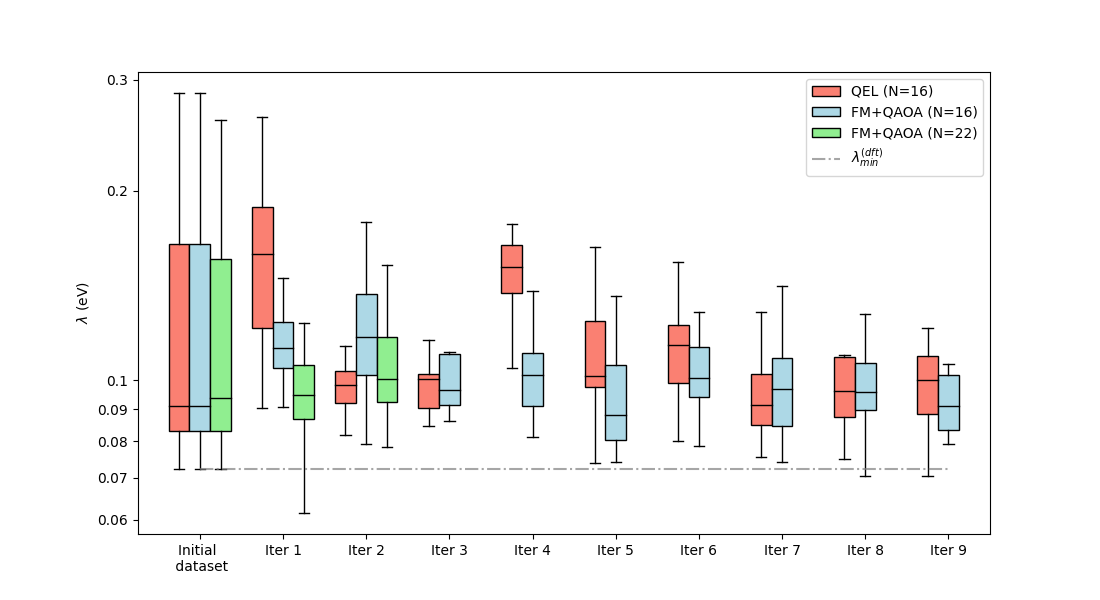}
    \caption{Box plot showing the distribution of the DFT-calculated hole reorganization energies $\lambda^{\rm (dft)}$ of the candidates predicted by each algorithm at every iteration of the Sequential Learning (SL) process. The box plot is drawn following the standard convention, with each box spanning from the first quartile to the third quartile of the data, so that the box length is equal to the inter-quartile range (the box width is arbitrary). The horizontal line inside the box marks the median, while the top and bottom whiskers represent the maximum and minimum values respectively. $\lambda^{\rm (dft)}_{\rm min}$ represents the lowest reorganization energy present in both the initial $N=16$ and $N=22$ datasets.} 
    \label{fig:box_plot}
\end{figure*}

We now move to benchmarking the algorithms with the larger datasets, starting with the $N=16$ dataset. 
Recall that this dataset contains the DFT-calculated hole reorganization energies $\lambda^{\rm (dft)}$ of 99 different chemical compounds represented by 927 bitstrings (approximately 9.4 duplicate bitstrings per compound). 
Since 16 bits are now used to encode the molecules, the complete bitstring solution space is now composed of all $2^{16}=65536$ possible bitstrings of length 16, significantly larger compared to the previous $N=7$ dataset.
Importantly, now we only have access to the $\lambda^{\rm (dft)}$ of a small fraction ($1.41\%$) of the solution space, meaning that we cannot check if a predicted candidate is the optimal solution because we do not know which is the lowest reorganization energy compound in the solution space.

Alternatively, we focus on whether the algorithms are able to predict structures with $\lambda^{\rm (dft)} <\lambda_{\rm{min}}^{\rm{(dft)}}$, where $\lambda_{\rm{min}}^{\rm{(dft)}}= 0.072$ eV is the lowest reorganization energy in both the initial $N=16$ and $N=22$ datasets. 
We denote any such compound as an \textit{extrapolative} compound. 
Finding an extrapolative compound means that the algorithm can successfully generalize to molecules in the low-$\lambda$ regime. 

\subsubsection{Hyperparameter tuning}

Following the same procedure as with the $N=7$ dataset, we performed a hyperparameter search to determine the optimal settings for the algorithms.
Regarding QEL, recall that for the $N=16$ and $N=22$ datasets we do not use a selection QNN, but rather a brute-force approach which does not have any hyperparameters. 
Therefore, only the depth of the learning QNN had to be tuned, for which a grid search was carried with depths ranging from 4 to 14.
The optimal depth was determined to be 8 layers by looking at the $R^2$ score on the isolated test averaged over 5 instances.
In the FM+QAOA, the depth of the QAOA circuit was fixed to 16 layers and the final state was sampled with $n_{\rm{shots}} = 10^5$ shots.  

\subsubsection{Results}

Figure \ref{fig:box_plot} depicts a box plot visualization of the DFT-calculated hole reorganization energy $\lambda^{\rm (dft)}$ of the selected candidates by each algorithm during the different iterations of the sequential learning process.
The FM+QAOA algorithm, when applied to the $N=16$ dataset, demonstrates a consistent trend of improvement, with the $\lambda^{\rm (dft)}$ of the predicted candidates increasingly clustering within the low-$\lambda$ regime (defined to be the region $<0.1$ eV).
On the other hand, QEL shows more variability, with some iterations failing to converge (e.g. iterations 1 and 4).
However, both algorithms are able to focus the predictions in the low-$\lambda$ regime by the later stages of the sequential learning process, especially after iteration 6, achieving similar final distributions.

\begin{figure}
    \centering
    \includegraphics[width=0.9\linewidth]{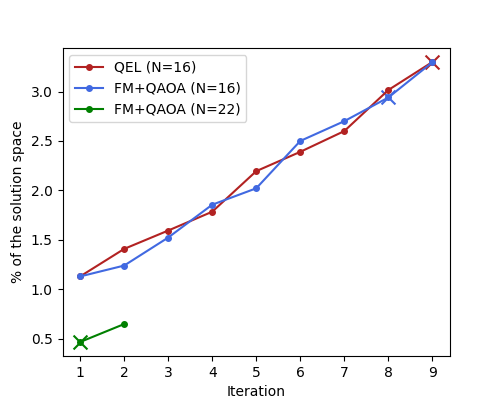}
    \caption{Percentage of the complete solution space used for training in each iteration. The cross in each line marks the point where extrapolation was first achieved.}
    \label{fig:n_data_points}
\end{figure}

It is illustrative to focus on the lowest reorganization energy compound predicted at each iteration, shown more clearly in the top plot of Figure \ref{fig:means_and_mins}.
Recall that we aim to predict chemical structures with a lower $\lambda$ than the minimum $\lambda$ in the initial dataset ($\lambda_{\rm{min}}^{\rm{(dft)}}=0.072$ eV), marked by the dashed black line in the plot. 
Both FM+QAOA and QEL closely follow each other predicting similar minimum $\lambda^{\rm (dft)}$, with the FM+QAOA performing slightly better in most iterations. 
In the first iterations, the lowest energy compounds have energies close to 0.085 eV. 
Once enough training data have been added through the sequential learning process (from iteration 4 onward), both algorithms are able to predict compounds with a minimum $\lambda^{\rm (dft)}$ close to 0.075 eV. 
Finally, the FM+QAOA algorithm predicts an extrapolative compound in iteration 8 using $2.94\%$ of the solution space for training. 
The predicted candidate has $\lambda^{\rm (dft)}$=0.071 eV, and its chemical structure is shown in Figure \ref{fig:best_compound_N16}. 
The QEL algorithm achieved extrapolation one iteration later, at iteration 9, and it predicted the same compound as the FM+QAOA with $\lambda^{\rm (dft)}=0.071$ eV. The QEL algorithm required a slightly larger training dataset than the FM+QAOA, using 3.33\% of the solution space for training.
Given that the sequential learning processes for both algorithms are completely independent to each other, it is remarkable that both QEL and FM+QAOA identified the same compound at almost the same iteration. 
Figure \ref{fig:n_data_points} shows the size of the training dataset in each iteration, relative to the total number of possible bitstrings in the solution space.

The bottom plot in Figure \ref{fig:means_and_mins} shows the mean $\lambda$ of the predicted candidates in each iteration.
Again, the FM+QAOA consistently decreases the mean $\lambda$ of the predictions, while in QEL it sharply increases in the iterations where it failed to converge. 
Nevertheless, both algorithms reach mean energies around 0.1 eV by the last few iterations, showing again a similar performance by the end of the sequential learning process.

\begin{figure}[tb]
    \centering
    \includegraphics[width=0.95\linewidth]{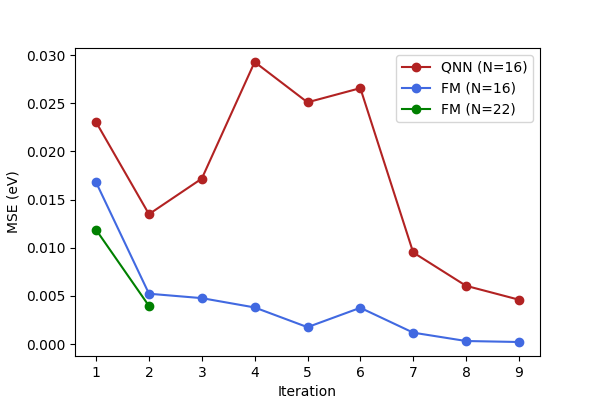}
    \caption{Mean Squared Error (MSE) of the predicted hole reorganization energy $\lambda^{\rm{(pred)}}$ of the candidate structures with respect to the DFT-calculated values $\lambda^{\rm{(dft)}}$. The Factorization Machine (FM) corresponds to the learning phase of the FM+QAOA algorithm and the Quantum Neural Network (QNN) to that of the QEL algorithm.}
    \label{fig:mse_accuracy}
\end{figure}

Up until now, we have discussed the overall performance of the algorithms based on the quality of the predicted candidates.
However, it is also interesting to solely focus on the performance of the models in the learning phase, as it constitutes a crucial step towards a working extremal learning algorithm.
Recall that the goal of the learning phase is to construct a model of the data that the selection algorithm can subsequently use to identify optimal candidates.
Figure \ref{fig:mse_accuracy} shows the Mean Squared Error (MSE) of the reorganization energy of the predicted candidates across iterations. In particular, it shows the MSE of the $\lambda^{\rm{(pred)}}$ predicted by the models in the learning phase with respect to the $\lambda^{\rm{(dft)}}$ values calculated using DFT for the candidate compounds. 
Recall that the FM+QAOA uses a Factorization Machine (FM) as a model in the learning phase, while QEL uses a Quantum Neural Network (QNN).
The FM consistently outperforms the QNN in terms of MSE across all iterations for the $N=16$ dataset. This improvement is especially pronounced in the mid-iterations (4 to 7), where QEL exhibits a noticeable spike in the MSE values.
Furthermore, the FM seems to consistently improve with each iteration, while the QEL algorithm again shows a more inconsistent progress, which is in line with our earlier observations.

\subsection{N=22 dataset}

We now move to the largest dataset that was investigated, constructed with a $N=22$ bits encoding. We recall that it contains 119 different heteroacene compounds represented by 24512 bitstrings (approximately 206 duplicate bitstrings per chemical compound).
In this case, only the FM+QAOA algorithm was trained with this dataset, as the computational resources to emulate QEL with 22 qubits were too large without drawing upon approximation methods such as tensor networks \cite{bidzhiev2023cloudondemandemulationquantum}, that were not considered in this work.
Given the large computational requirements, the depth of the QAOA circuit was limited to 14 layers, and $n_{\rm{shots}} = 10^7$ where used to sample the final QAOA state.

The FM+QAOA performed very well with the 22 bits dataset, predicting octacene as an extrapolative compound with $\lambda^{\rm (dft)}$=0.062 eV already at the first iteration (see Figures \ref{fig:means_and_mins} and \ref{fig:box_plot}).
This was achieved using 19873 bitstrings for training, which represents 0.47\% of the bitstring solution space. 
Even though the absolute number of bitstrings used for training to achieve extrapolation is larger than in the $N=16$, it represents a significant reduction in terms of relative size to the total number of bitstrings, as the FM+QAOA required 2.94\% and QEL 3.33\% to predict an extrapolative design with $N=16$.
Therefore, we can observe that as the solution space grows, the absolute amount of data needed for extrapolation increases, but the fraction of the solution space needed for effective learning decreases.

The structure of the predicted extrapolative molecule was also very interesting (linear structure with no 5-membered rings), and we note that such structures have also been identified as a low-$\lambda$ structure in a previous study \cite{Marques2021}.
One more iteration was done after achieving extrapolation at iteration 1, which did not provide any compounds with lower $\lambda^{\rm (dft)}$ despite improving the mean $\lambda^{\rm (dft)}$ of the predicted compounds (bottom plot of Figure \ref{fig:means_and_mins}).

\subsection{Chemical insights}

\begin{figure}[tb]
    \centering
    \includegraphics[width=0.85\linewidth]{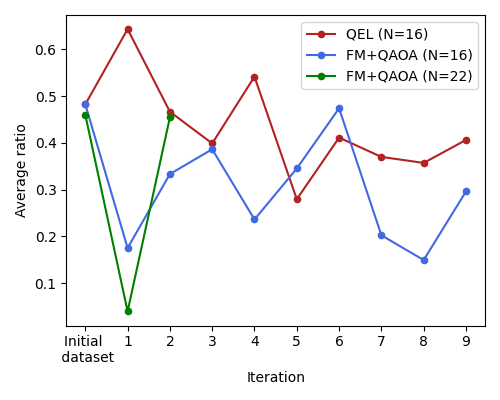}
    \caption{Average ratio of 5-membered rings in the structures predicted by each algorithm.}
    \label{fig:average_5_members}
\end{figure}

\begin{figure*}[tb]
    \centering
    \subfloat[$\lambda^{\rm (dft)}=0.071$ eV]{%
        \includegraphics[width=0.19\textwidth]{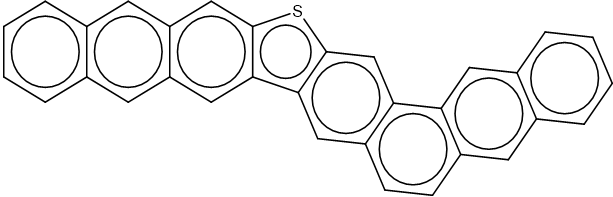}
        \label{fig:best_compound_N16}
    }
    \hfill
    \subfloat[$\lambda^{\rm (dft)}=0.062$ eV]{%
        \includegraphics[width=0.19\textwidth]{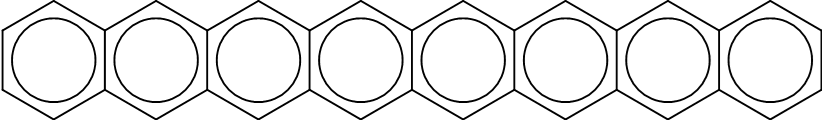}
        \label{fig:best_compound_linear}
    }
    \hfill
    \subfloat[$\lambda^{\rm (dft)}=0.075$ eV]{%
        \includegraphics[width=0.19\textwidth]{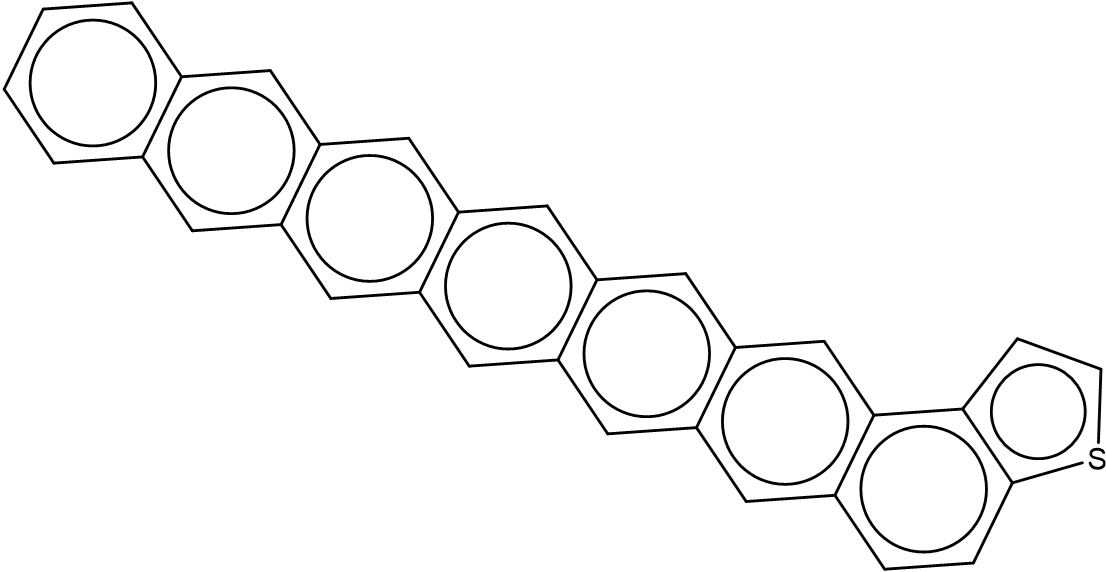}
        \label{fig:best_compound_semilinear_1}
    }
    \hfill
    \subfloat[$\lambda^{\rm (dft)}=0.075$ eV]{%
        \includegraphics[width=0.19\textwidth]{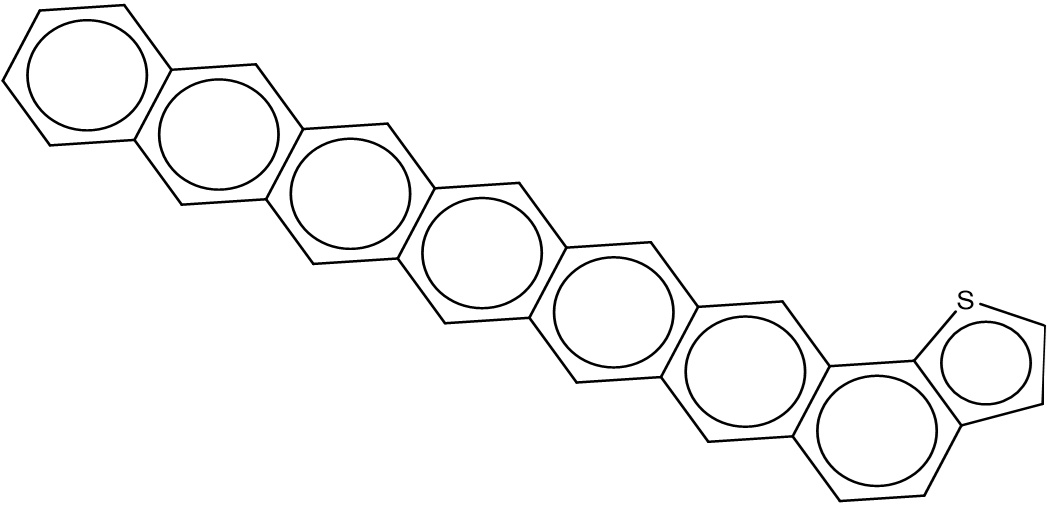}
        \label{fig:best_compound_semilinear_2}
    }\hfill
    \subfloat[$\lambda^{\rm (dft)}=0.072$ eV]{%
        \includegraphics[width=0.20\textwidth]{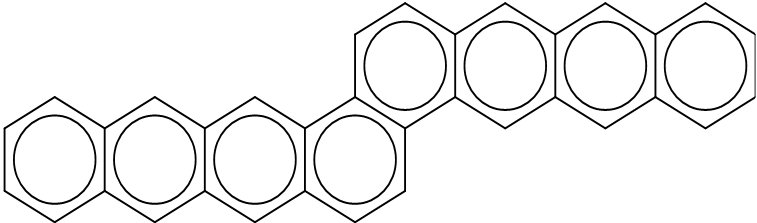}
        \label{fig:best_compound_semilinear_alternative}
    }

    \caption{Different heteroacene structures and their DFT-calculated hole reorganization energies $\lambda^{\rm (dft)}$. (a) Best compound predicted by training with the $N=16$ dataset, found in iteration 8 of FM+QAOA and iteration 9 of QEL. (b) Octacene was the best compound predicted with the $N=22$ dataset, found in the first iteration of the FM+QAOA algorithm. It has a completely linear structure and no heteroatoms (all rings are benzene rings). Figures (c) and (d) show two structures present in the initial $N=22$ dataset which are equal to (b) in all but the last ring, which is a thiophene instead of a benzene. Lastly, (e) shows the low-$\lambda$ compound predicted by the FM+QAOA algorithm with the $N=22$ dataset in the alternative first iteration when compounds (c) and (d) where removed.}
    \label{fig:chemical_structures_diagram}
\end{figure*}

Having discussed the algorithmic performance, we now dedicate the rest of the Section to the chemical insights arising from the compounds predicted by the quantum-enhanced algorithms. 

Figure \ref{fig:chemical_structures_diagram} shows different hetereoacene structures that were predicted by or used for training in the algorithms, together with their DFT-calculated hole reorganization energies.
The compound in Figure \ref{fig:best_compound_N16} is the extrapolative design predicted by the FM+QAOA and QEL algorithms with the $N=16$ dataset at iterations 8 and 9, respectively. 
In Figure \ref{fig:best_compound_linear} the extrapolative compound (octacene) predicted by the FM+QAOA with the $N=22$ dataset is presented. 
There are two properties worth noting in this heteroacene structure. 
Firstly, we observe that the structure is completely linear, as all rings in the chain are connected along a straight line.  
Secondly, none of the rings contain heteratoms, so that all rings are 6-membered.
The latter aligns well with previous research, where a low ratio of 5-membered rings in heteroacene structures has been linked to low hole reorganization energies with the most preferable range of the ratio being 0.2-0.4 in terms of minimizing the reorganization energy \cite{Matsuzawa2020}.
In fact, Figure \ref{fig:average_5_members} shows how as the sequential learning process advances, the ratio of 5-membered rings in the predicted candidate compounds also tends to decrease. 
This is in line with our previous reasoning because as iterations pass, the $\lambda$ of the predicted candidates also decreases, due to the training dataset improving in both size and quality. 

It is remarkable that the FM+QAOA algorithm was able to extrapolate to the low energy compound in Figure \ref{fig:best_compound_linear} with the $N=22$ dataset already in the first iteration, using only 0.47\% of the bitstring solution space for training. 
Upon further examination of the training dataset, we found two compounds with a very similar structure to \ref{fig:best_compound_linear} which also have low reorganization energy (both with $\lambda^{\rm (dft)}=0.075$ eV). 
These two compounds, shown in Figures \ref{fig:best_compound_semilinear_1} and \ref{fig:best_compound_semilinear_2}, are equal to octacene in all but the last ring, where a thiophene ring is present instead of a benzene ring, with the sulfur heteroatom placed differently in each of them. 
Also, the compounds are "semi-linear" in the sense that only the last ring containing the heteroatom is not oriented following the line defined by the rest of the rings.

In order to check the effect of the two semi-linear compounds on the algorithm predicting the extrapolative compound, we run the first iteration of the FM+QAOA algorithm again but removing the two semi-linear compounds from the dataset. 
As it turned out, the FM+QAOA algorithm was still able to predict the extrapolative compound \ref{fig:best_compound_linear}. 
In fact, another compound with low reorganization energy $\lambda^{\rm (dft)}=0.072$ eV was predicted which also had a semi-linear structure with  all rings being 6-membered (benzenes), shown in Figure \ref{fig:best_compound_semilinear_alternative}. 
In conclusion, the results of this alternative first iteration were similar to the original run, with 21 out of the 50 best selected bitstrings overlapping, indicating that the presence of the two semi-linear compounds was not the main cause behind the FM+QAOA algorithm finding the extrapolative compound. 
We should keep in mind that we are analyzing the results of a single iteration of the sequential learning process, and therefore a lot of variability can be expected, especially given the vast size of the solution space with $N=22$ bits. This means that no strong conclusions can be derived from the fact that the algorithm performed slightly better without the two semi-linear compounds.

\section{Conclusions}
\label{sec:conclusions}

In this work, we have explored the use of two quantum-enhanced algorithms for a materials discovery task.
Specifically, we aimed to predict heteroacene structures with low hole reorganization energy $\lambda$, which is a reliable estimator of the carrier mobility in organic semiconductors.
The algorithms employed here follow the extremal learning framework, which consists of a learning phase where a ML model is trained to fit the data and a selection phase where the optimal (or extremal) candidates are selected.
The algorithms are quantum-enhanced because in each of them a different phase is done via a quantum algorithm: in QEL, the learning phase is quantum while the selection phase is classical, whereas in FM+QAOA, the learning phase is classical and the selection phase is quantum.

Our numerical experiments showed that both algorithms were able to predict heteroacene structures with lower hole reorganization energy than the lowest energy compound present in the initial dataset ($\lambda^{\rm (dft)}<\lambda^{\rm{(dft)}}_{\rm{min}}=0.072$ eV), which we denote as an extrapolative compounds, proving good generalization capabilities to the low-$\lambda$ regime. 
When trained with the $N=16$ dataset, the FM+QAOA and QEL algorithms predicted an extrapolative compound with $\lambda^{\rm (dft)}=0.071$ eV after 8 and 9 iterations of the sequential learning process, respectively. 
Both achieved a similar performance in terms of the lowest predicted reorganization energies, although the QEL algorithm offered less consistent predictions and less accurate models in the learning phase.
Furthermore, the FM+QAOA was also applied to a larger dataset with $N=22$, predicting a compound (octacene) with $\lambda^{\rm (dft)}=0.062$ eV already in the first iteration while using a smaller portion of the solution space for training compared to the $N=16$ case.
Crucially, the previous extrapolative structures were suggested whilst working on samples of a classically curated dataset deriving from a massive theoretical screening of $\sim 7$M molecules and a record $\lambda^{\rm{(dft)}}_{\rm{min}}\sim0.055$ eV. This observation makes the final results presented in this paper with few iterations and a much smaller dataset far from trivial. 
These results also suggest that improved performance can be achieved by increasing the number of bits $N$ in the encoding, paving the way to the application of such algorithms to datasets encoding a larger chemical space.

The predicted low-$\lambda$ compounds offered interesting chemical insights, as they presented two key common features: a linear or semi-linear structure and a majority of 6-membered rings. This is in line with previous research where it has been shown that such structures tend to show low values of $\lambda$ \cite{Matsuzawa2020, Marques2021}. For example, octacene is found to be the lowest $\lambda$ compound in our study, whereas in a previous study, nonacene is listed as one of the lowest-$\lambda$ compounds that were de novo designed \cite{Marques2021}. 

In conclusion, this work aimed to explore quantum-enhanced extremal learning algorithms in realistic, state-of-art scenarios of materials discovery, focusing on variational algorithms amenable to implementation in near-term quantum processors.
The numerical results presented in this work were computed via noiseless emulation on large classical computing resources. In order to unlock the benefits from the quantum stage of our algorithms, further work studying the scalability and quantum-hardware implementation must be carried. 
Future work could focus on improving the scaling of the proposed algorithms with the size of the datasets. 
For example, in QEL, the brute-force selection could be substituted with a QNN adapted to extremize over discrete inputs, or the training process of the QAOA algorithm could be subject to improvements to benefit convergence in larger circuits. 
Our results, although necessarily limited, offer an exciting prospect regarding the use of Quantum Machine Learning for material discovery.

\begin{acknowledgements}
IFG and SV thank Antonio A. Gentile, Arthur M. Faria, Anton Quelle and Jaap Kautz for useful discussions and valuable feedback, Sylvain Maret for his help administering the HPC resources, Daniel Guijo for his project management skills and Vincent E. Elfving for his initial help setting up the project. 
\end{acknowledgements}

\bibliographystyle{IEEEtran}
\bibliography{bib}

\end{document}